\documentclass[runningheads]{llncs}
\usepackage[T1]{fontenc}
\usepackage{graphicx}
\graphicspath{{figures/}}
\usepackage{color}

\usepackage{amsmath}
\usepackage{amsfonts}
\usepackage{comment}
\usepackage{enumitem}
\usepackage{listings}
\usepackage{longtable}
\usepackage{multirow}
\usepackage{subcaption}
\usepackage[notransparent]{svg}
\usepackage{adjustbox}
\usepackage{pifont}

\usepackage[hidelinks]{hyperref}

\begin{document}

\title{Understanding Layered Portability from HPC to Cloud in Containerized Environments}
\titlerunning{Characterizing Container Portability from HPC systems to the Cloud}
\author{Daniel Medeiros \and Gabin Schieffer \and Jacob Wahlgren \and Ivy Peng}
%
%
\institute{Department of Computer Science,\\ KTH Royal Institute of Technology, Sweden\\\email{\{dadm, gabins, jacobwah, ivybopeng\}@kth.se}\\}

\maketitle              
\begin{abstract}

Recent development in lightweight OS-level virtualization, containers, provides a potential solution for running HPC applications on the cloud platform. In this work, we focus on the impact of different layers in a containerized environment when migrating HPC containers from a dedicated HPC system to a cloud platform. On three ARM-based platforms, including the latest Nvidia Grace CPU, we use six representative HPC applications to characterize the impact of container virtualization, host OS and kernel, and rootless and privileged container execution. Our results indicate less than 4\% container overhead in DGEMM, miniMD, and XSBench, but 8\%-10\% overhead in FFT, HPCG, and Hypre. We also show that changing between the container execution modes results in negligible performance differences in the six applications.

\keywords{Cloud and HPC Convergence \and Containers \and ARM \and Performance}
\end{abstract}

\section{Introduction}
\label{sec:intro}

High accessibility to a wide variety of computing resources, timely access to new hardware, and cost-effectiveness motivate running HPC applications in the cloud, moving towards the convergence of HPC and Cloud~\cite{araujo2023libcos,de_sensi_noise_2022,medeiros2023kub,reed_hpc_2023,milroy_one_2022,medeiros2023gpu}. Today, lightweight containers are replacing virtual machines (VM) to become the widely used virtualization and isolation mechanism on the cloud. Running HPC applications in a containerized environment is one main distinction from running them in bare metal on-premise HPC systems. Popular container engines on HPC systems, such as Podman~\cite{gantikow2020rootless}, Singularity~\cite{kurtzer2017singularity} and Charliecloud~\cite{priedhorsky2017charliecloud}, are specially designed for \texttt{rootless}/unprivileged container execution while Docker, the de-facto solution on the cloud, runs containers through a \texttt{root-owned} daemon. On the convergence of HPC and cloud, one likely scenario is that a user builds an image of an HPC application from an HPC environment and deploys it on instances in a cloud environment. Therefore, the interoperability across the two environments and the associated performance impact from each layer in the dependency chain is important but not fully explored. However, despite extensive studies that have characterized and optimized the performance of HPC applications in a virtualized environment, few works have explored the portability and associated performance loss in containerized HPC applications moving between HPC and the cloud platforms~\cite{keller2023containers}.

The portability of containerized applications ensures that an image can be built on a host in one infrastructure and deployed on another host, likely in another infrastructure. As illustrated in Figure~\ref{fig:hierarchy}, hardware, OS and kernel, and container engine are the three major components likely different in the two environments. Within a container, its layered architecture may also result in changes in one dependency layer cascading to subsequent layers. Although interoperability ensures that a containerized application can execute across different platforms, the combination of software in use in each layer may impact performance differently. For instance, the kernel of the host OS will be used by the containerized application, and it may consist of different versions on the building and the deployment platforms. 

Besides the difference in software stacks, ARM-based processors are commonly used and offered on the cloud. Modern cloud-enhanced ARM-based processors provide extensions for security, efficient virtualization, and lower energy consumption than their x86 counterparts. However, in the HPC landscape, Fugaku is the only ARM-based supercomputer to reach the top 500 ranking during June 2020 and May 2022, while other top supercomputers are x86 based on either Intel or AMD processors.

In this work, we evaluate the impact of the differences in three layers in containerized HPC applications on the build and deployment platforms. We use six HPC applications to represent diverse CPU or Memory-intensive workloads. First, we evaluate the overhead of popular container engines in the cloud and HPC on three generations of ARM processors, including the latest Nvidia Grace Processor. Second, we evaluate the impact of changing the "OS layer" from the image when moving from the build platform to the deployment platform. Finally, the impact of switching between rootless and root-owned container engines on the deployment platform is evaluated. In summary, we made the following contributions:
\begin{itemize}[noitemsep,topsep=0pt]
    \item We quantify the containerization overhead in six HPC applications on three ARM processors including the Nvidia Grace CPU.
    \item We evaluate the impact of the changed OS layer in Docker and Podman when moving from the build to deployment platforms.
    \item We evaluate the impact of switching between rootless and root privileges when moving from the build to deployment platforms.
\end{itemize}


\section{Background}
\label{sec:bg}

\begin{figure}
    \centering
    \includegraphics[width=0.5\linewidth]{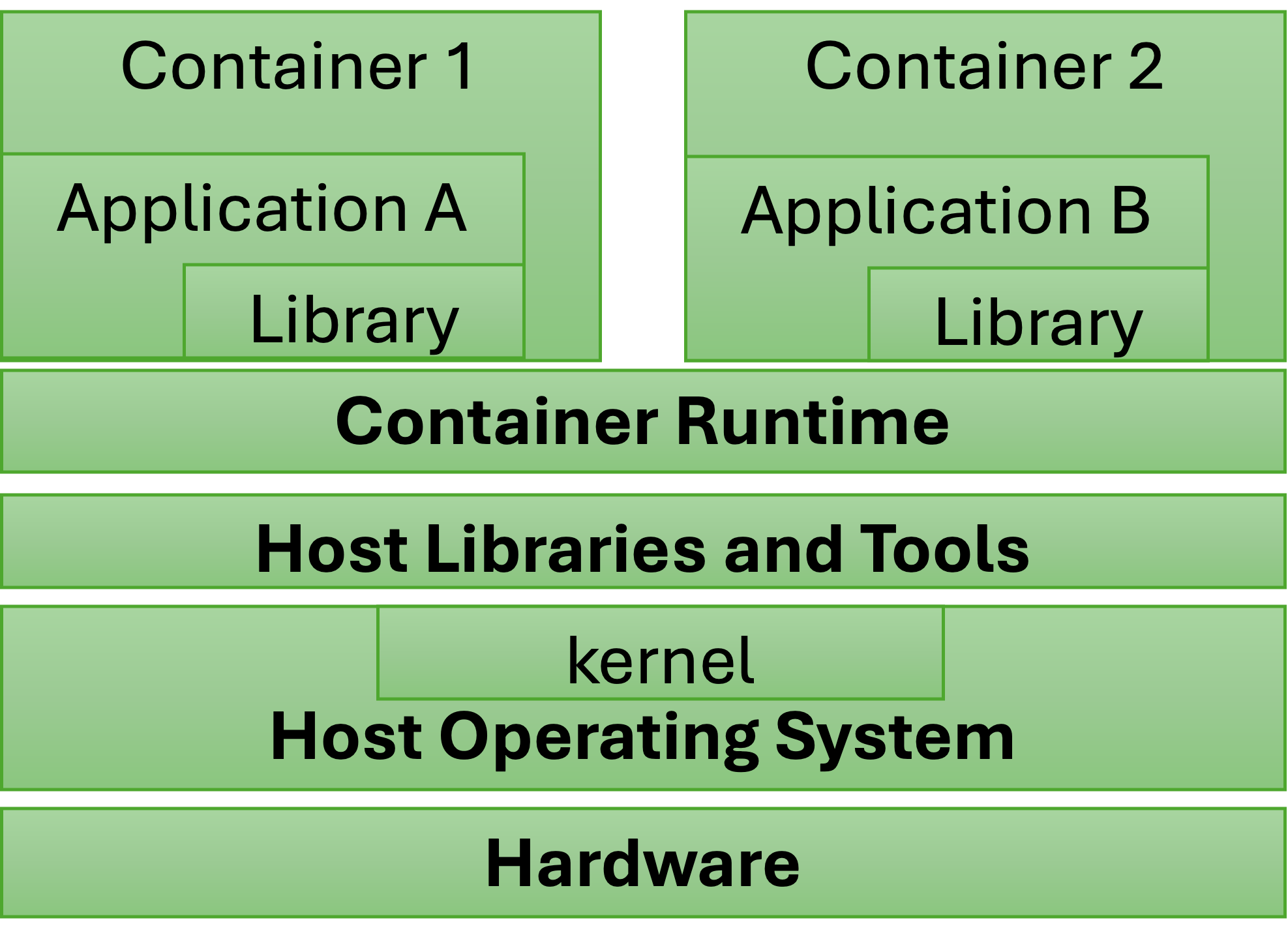}
    \caption{A hierarchical view of different layers in a containerized environment}
    \label{fig:hierarchy}
\end{figure}

In this work, we focus on identifying the impact of different layers in a containerized environment when migrating an HPC container between dedicated HPC systems and the cloud. In Figure~\ref{fig:hierarchy}, we illustrate the main layers that may change when moving from one system to another. Here the hardware layer is the lowest level and it fundamentally impacts the obtainable performance from a platform. A container shares the same Linux kernel as the host OS, but might have a different root file system and toolchain associated with the image it is running that may affect performance: the image, for example, could use \texttt{musl libc} as runtime while the host OS could use \texttt{glibc}, or different versions of compilers when building code. Finally, although standard container runtime and image specifications ensure usability from one platform to another, the specific container engine in use and its container execution mode on a host may impact performance differently.   

\begin{figure}
    \centering
    \includegraphics[width=0.75\linewidth]{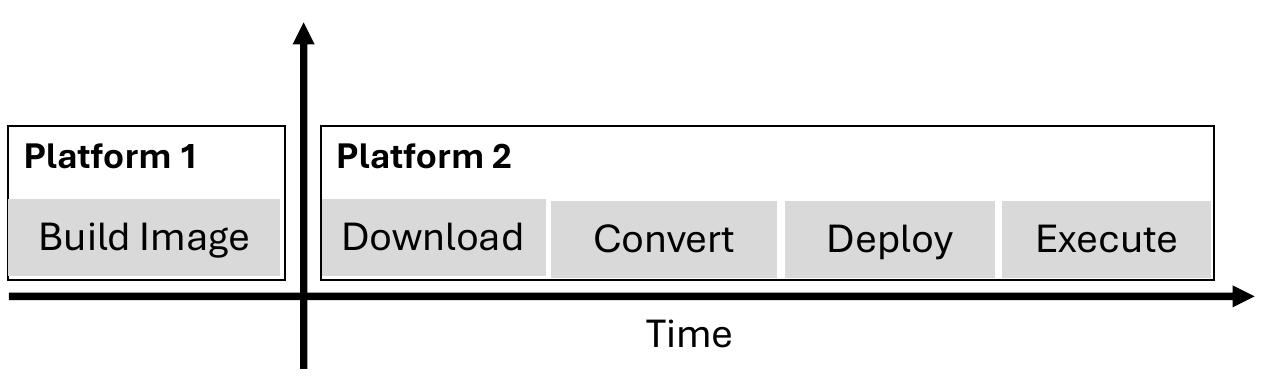}
    \caption{The main stages of building a image on a platform and deploying it on another platform.}
    \label{fig:stages}
\end{figure}

\textbf{Container Engine} provides OS-level virtualization and isolation through the kernel's \texttt{namespaces} and \texttt{cgroup} in Linux. Containers are a lightweight way for managing and deploying microservices in the cloud as it does not virtualize a full guest OS for each instance of a service as required by hypervisors. On HPC systems, \textit{rootless} container engines are popular, such as Singularity~\cite{kurtzer2017singularity}, Podman~\cite{gantikow2020rootless}, while Docker~\cite{merkel2014docker} is the de-facto solution on the cloud platform. Podman has syntax compatible with Docker but does not require a privileged daemon to be running all the time as in Docker. In this work, we tackle an emerging scenario where HPC applications are migrating between native HPC systems and the cloud in portable containers. Figure~\ref{fig:stages} illustrates the main stages and platforms involved in the migration, on the target platform. There, a previously built image is downloaded and stored, converted if necessary (i.e., if an image is built on Docker and will be executed by Podman), a container is created and deployed, and then the application is ultimately executed inside the container. 

\textbf{Host-specific Libraries and Tools} HPC applications have high requirements to get near-hardware native performance from high-end and specialized hardware units. Vendors often provide libraries and tools that are specifically tuned to the hardware on the host platform.

\textbf{Host-specific Libraries and Tools} are often provided by vendors and are specifically optimized for the hardware on the host platform. These tools are crucial for achieving near-native performance of HPC applications, given the stringent requirements of high-performance computing environments. For instance, when HPC applications in a container can link to the host's MPI implementation with hardware-specific optimizations, they can mitigate the performance loss due to inefficient usage of network and interconnects. Profiling tools are critical for diagnosing and optimizing applications. As processor-specific hardware performance counters often provide insights into the utilization of specific units, it is also important for HPC applications to link to the host's hardware-specific profiling capabilities.

\textbf{Hardware} often differentiates in Cloud and HPC environments as they deploy processors tailored for their workloads of priority. For instance, HPC platforms only use high-end multi-core processors with high computing power to meet the needs of HPC applications and x86 architecture is dominantly in use. In the Cloud environment, a high variety of processors from medium to high-end, can be chosen to different users' target on cost and performance. In recent years, ARM processors have gained increasing popularity on the cloud, represented by AWS' Graviton3 and the Ampere Altra processors. 
\section{Methodology}
\label{sec:method}
In this work, we characterize the impact of different layers in a containerized environment when migrating between the HPC system and the cloud. We identify three common changes when migrating an HPC container to the cloud environment. First, applications on HPC systems often run in bare metal, while on the cloud platform, applications are running within virtualization, and application users need to understand the overhead: not only on the main computational phase but also the deployment phase for creating and setting up a container. 

The second likely change to applications that are used in dedicated HPC systems comes from the kernel of host OS and host-specific libraries, which are shared by all containers, on a deployment platform. The most common libraries used in HPC applications are the MPI library and math libraries like BLAS and FFT. Vendors often provide highly optimized libraries for a specific architecture on an HPC system. Finally, when an HPC container migrates to the cloud, unlike the rootless container execution in HPC systems, it can be executed as a privileged container execution on the cloud. Due to the difference in resource isolation mechanisms, the two container execution modes may impact applications' performance if they require specific kernel-level services.

To understand the above changes quantitatively, we design a set of experiments based on the layered architecture of a container. In particular, we control the software in use in all layers as introduced in Section~\ref{sec:bg} and only change a specific layer to isolate its impact. To ensure wide coverage of the study, we perform comparative studies on three platforms that feature different hardware, OS, host-specific libraries, and container engines. We also select a set of six applications with different characteristics in compute, memory access, and communication, to represent the diverse HPC workloads on HPC systems.

We first quantify the overhead of container-based virtualization. For this, we ensure that each software layer inside a container is identical to that on a platform in bare metal. Second, we use two different platforms to emulate the scenarios of changing OS kernel and host libraries layers (base image) in a containerized environment. Finally, we run HPC containers in both the rootless and privileged modes in Docker and Podman and compare the performance difference between the two modes of the same container runtime on a set of HPC applications. Table~\ref{tab:layers} summarizes the different options evaluated in each layer in this work.

\begin{table}[bt]
    \centering
    \caption{A summary of evaluated layers on the three containerized environments}
    \label{tab:layers}
    \begin{tabular}{|c|c|c|}
    \hline
    Container Engine & Docker | Podman \\\hline
    Execution Mode & rootless | root \\\hline
    App Libraries &  BLAS | MPI | FFT \\\hline
    OS & RHEL | Rocky Linux | Ubuntu | OpenSUSE \\\hline
    Kernel & Linux \\\hline
    Processor & Nvidia Grace CPU | ARM Ampere | APM X-GENE \\\hline
    Core & Neoverse N2 | Neoverse N1 | Cortex-A57 \\\hline    
    \end{tabular}
\end{table}

\textbf{Hardware Infrastructure}
In this work, we use three different platforms to evaluate our workloads, namely CloudLab, Pilot and Sleipner.

The Sleipner platform has the latest Nvidia Grace CPU. The node has 72 64-bit Neoverse V2 cores running at 3.4 GHz and 592 GB DRAM. The Pilot platform is based on the Altera Altra processor. Each compute node has 80 64-bit Neoverse N1 cores running at 3.0 GHz and is equipped with 526 GB DRAM and 2.4TB SSD. On CloudLab, we used the compute node with an APM X-GENE (Cortex-A57) CPU, including eight 64-bit Atlas/A57 cores running at 2.4 GHz, 64GB ECC DRAM, and 120 GB of flash memory.

\begin{table}[bt]
\centering
\caption{A summary of major software layers used in Bare Metal and Container on the three computing platforms}
\label{tab:software-layers}
\resizebox{\columnwidth}{!}{%
\begin{tabular}{|l|l|l|l|l|l|l|}
\hline
 &\multicolumn{2}{|c|}{Sleipner} &\multicolumn{2}{|c|}{Pilot} &\multicolumn{2}{|c|}{CloudLab}\\\hline
 & \textbf{Host} &\textbf{Container} & \textbf{Host} &\textbf{Container} &\textbf{Host} &\textbf{Container} \\\hline
\textbf{OS}   & RHEL 9.3 & Rocky Linux 9.3 & OpenSUSE 15.5 &OpenSUSE 15.5 & Ubuntu 22.04 &Ubuntu 22.04\\\hline
\textbf{GCC}  & GCC 11.4 & GCC 11.4 & GCC 13.2 & GCC 13.2 & GCC 11.4 & GCC 11.4    \\\hline
\textbf{MPI}  & Open MPI 5.0.2 & Open MPI 5.0.2& Open MPI 5.0.2 & Open MPI 5.0.2 & Open MPI 5.0.2 & Open MPI 5.0.2\\\hline
\textbf{BLAS} & OpenBLAS 0.3.26 & OpenBLAS 0.3.26 & OpenBLAS 0.3.26& OpenBLAS 0.3.26& OpenBLAS 0.3.26& OpenBLAS 0.3.26\\\hline
\end{tabular}
}
\end{table}

\textbf{Software layers} mainly consists of the stack of OS and kernel, host libraries and compilers in either bare metal or a container environment. Table~\ref{tab:software-layers} specifies the main software layers in Bare Metal and Container on the three platforms. The Sleipner platform is running Red Hat Enterprise Linux 9.3 (Kernel: 5.14.0), GCC v11.5.1 and Docker 25.0. The software stack used in the Pilot platform includes OpenSUSE 15.5 (Kernel 5.14.21) and Podman 4.7.2. while the CloudLab platform is running Ubuntu 22.04, Linux Kernel 5.14.21 and Docker 20.10.

\subsection{Applications}
In this work, we use six applications representative of diverse HPC workloads and algorithms. For each application, two input sizes were selected, one meant to be used at the Sleipner and Pilot platforms, and the other to be used at the Cloudlab platform. The reason for different workloads is that the Cloudlab platform is significantly older and slower than the other two, hence a similar workload input would take much longer to execute there. 

\textbf{DGEMM} performs a double-precision matrix-matrix multiplication in the format $C = \alpha A *  B + \beta C$, where A, B and C are matrices and $\alpha$ and $\beta$ are scalars. We tested with two input sizes of $5000 \times 5000$ and $20000 \times 20000$, and five as the number of repetitions within each trial. The DGEMM benchmark we used\footnote{Crossroads Benchmark: \url{https://www.lanl.gov/projects/crossroads/benchmarks-performance-analysis.php}} is multi-threaded through OpenMP and has no MPI support.

\textbf{Hypre}~\cite{falgout2002hypre} is a library for high-performance preconditioners and solvers that help solve sparse and linear systems of equations. Hypre uses MPI as a programming model. We used the provided example of solving the convection-reaction-diffusion problem. We tested with two input problems of grid size (n) of 300 and 3000. 

\textbf{XSBench} \cite{tramm2014xsbench} is a proxy app of a Monte Carlo neutron transport application. The application has memory-bound characteristics. We built XSBench with MPI support and tested it with the built-in small and large problem sizes. 

\textbf{HPCG} \cite{dongarra2015hpcg} is a Performance Conjugate Gradients benchmark that measures the performance of a system through the usage of basic operations, which includes sparse matrix-vector multiplication, global dot products, among others. We used input matrices of dimension $128 \times 128 \times 128$ and $256 \times 256 \times 256$. 

\textbf{MiniMD} is a parallel molecular dynamics mini application. It has a similar algorithm as LAMMPS (Lennard-Jones/EAM system) but has a simpler and smaller code (5000 lines vs over 200000 in the latter). Two input problems of size $64 \times 64 \times 64$ and $256 \times 256 \times 256$ are used. 

\textbf{fftMPI} \cite{plimpton2018fftmpi} is a library used to perform 2D or 3D Fast Fourier Transforms in parallel. Compared to 1D FFT, 2D and 3D 3D FFT also include memory-bound transposes. FFT is widely used in scientific applications and signal processing. Two grids are used for evaluation: $300 \times 300 \times 300$ and $400 \times 400 \times 400$. As its names implies, fftMPI uses MPI for calculations. 


\textbf{Profiling}. We use Linux's \texttt{perf} tool to collect hardware performance counters on data locality, memory access, and instruction pipeline. In particular, we needed to explicitly give hardware access to Docker through the usage of the feature \texttt{CAP\_SYS\_ADMIN} in the Sleipner platform.

\section{Evaluation}
In this section, we discuss three major questions regarding the usage of containers in HPC workloads. When moving towards the convergence of HPC and Cloud, one common scenario is to build the image on one platform and deploy it on another platform on the cloud. 

Unless said otherwise, the execution times displayed here are the ones reported by the application (with the exception of XSBench, which does not report by default).  

\subsection{How much impact does the usage of a containerized environment for application deployment cause?}\label{contimp}
\begin{figure}[ht]
    \includegraphics[width=\textwidth]{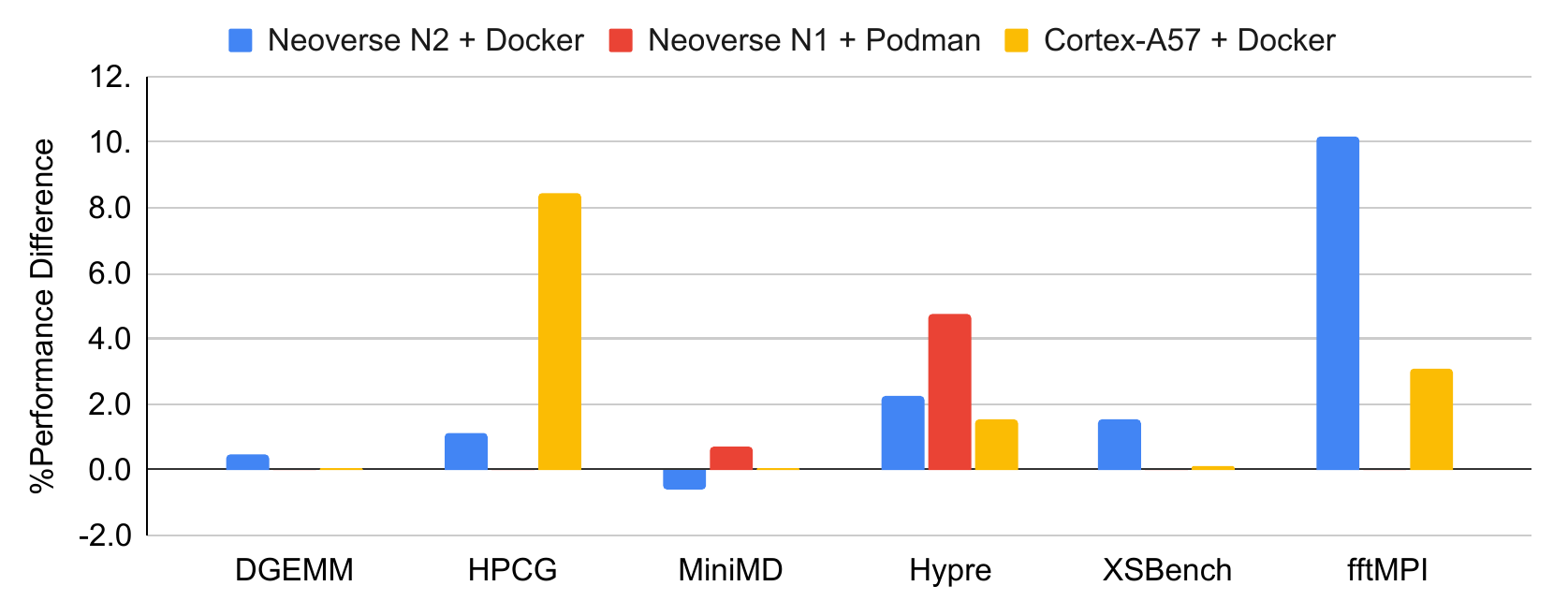}
    \caption{Results for the six HPC applications running on three generations of ARM processor, including Nvidia Grace, ARM Ampere Altra, and APM X-GENE.}
    \label{fig:bare-container}
\end{figure}

The first question deals with the overall performance of the usage of containers. Given a very similar stack between the Host OS and the Container OS, how much is the impact generated only by the isolation of process/namespaces caused by containers in terms of application execution time? Is this impact consistent in multiple ARM-based architectures? 

To answer this, we designed an experiment where, in three different ARM processors, we use the same Host OS, the same Container OS for each machine and also the same software stack - this includes compilers and application libraries/dependencies. We believe that this would enable a fair comparison between the two environments. Table \ref{tab:software-layers} lists the software environment in each system. 

Each of the six applications listed in Section \ref{sec:method} was executed at least three times, and the plots with the performance difference data can be seen in Figure \ref{fig:bare-container}. The overhead is less than 2\% for non-MPI applications (DGEMM and HPCG) in the more recent platforms. Among the MPI-based applications, fftMPI is the clear outlier where the performance difference to bare metal is about 10\%, while the other applications usually have less than 4\%. Given that the authors were not able to execute fftMPI using the entire CPU resources of Sleipner or Pilot in a containeirized environment, but managed to do so on bare metal, we consider that there might be some interaction with the resource isolation provided by the containers and the application itself that might be causing the slowdown.

\begin{figure}[bt]
    \includegraphics[width=\textwidth]{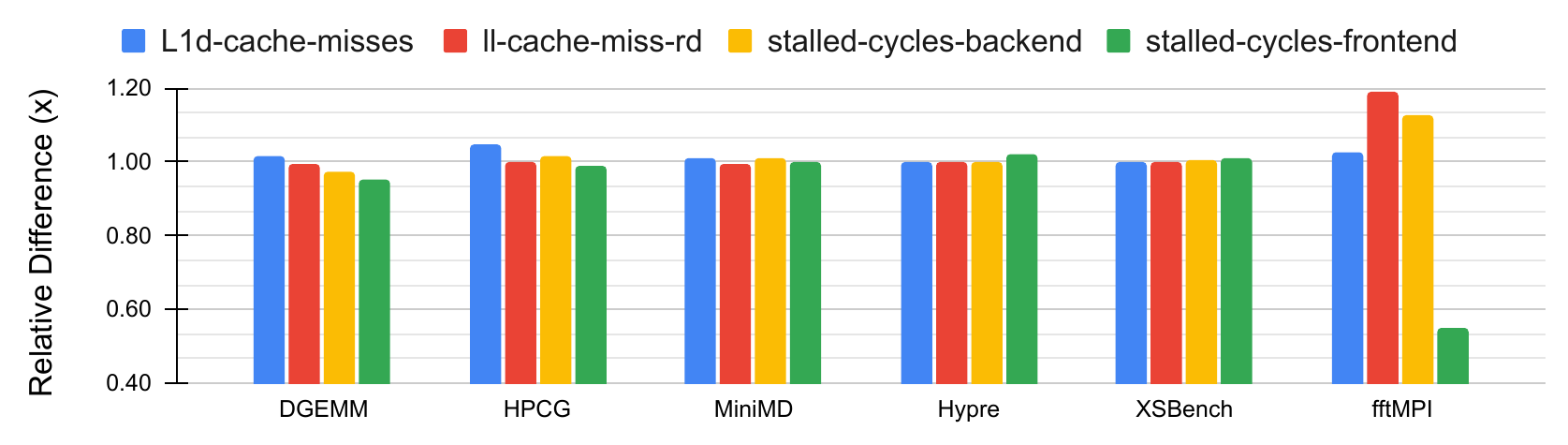}
    \caption{The relative changes in four performance counters in six HPC applications on the Sleipner platform (Nvidia Grace) in Docker container and Bare Metal.}
    \label{fig:performance-counter}
\end{figure}

Furthermore, on the Sleipner platform (the Nvidia Grace processor), we further analysed performance counters that are highly related to data locality, memory access, and instruction execution pipeline, as specified in Section~\ref{sec:method}.
Figure~\ref{fig:performance-counter} displays the relative changes in the performance counters and, as the impact in performance execution is minimal, the results displayed over there also show very minor changes. The results show that in absolute values, we observe that the ones pertaining to the stalls and cache groups had a direct correlation with the execution time, while the SVE group didn't change significantly in its values. 

\begin{figure}[bt]
    \includegraphics[width=\textwidth]{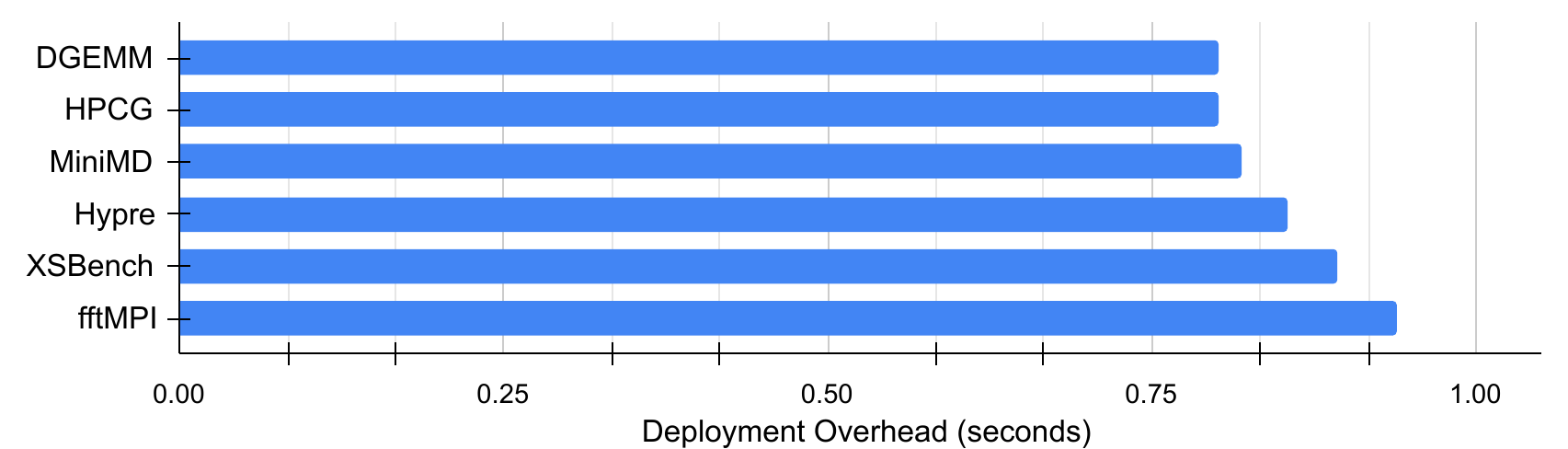}
    \caption{The overhead of Docker container deployment on the Nvidia Grace-based platform in the six applications.}
    \label{fig:deployment-overhead}
\end{figure}

Finally, another usual point when handling performance questions would be how long the Docker application itself takes to start the container. We measure both the time to execute the application using the \texttt{time} command, so this would consider the application start-up time, as well as the total time for the end-to-end process. In this case, we built an 8-layered image that amounted to roughly 1 GB in size and contained all necessary dependencies to run all the showcased applications. Figure \ref{fig:deployment-overhead} shows the collected data with the difference between such times, and overall displays that the absolute time takes usually less than 1 second, which is irrelevant if the application runs for a long time. That said, we note that it is a common practice within this field to produce very small images in a step called "multi-stage build" where a second image contains only the minimum necessary to run a certain application (the image for Alpine Linux, for example, has around 30 MB), while the compilation happens somewhere else, and we would expect this to generate even less overhead.

\subsection{How much does using a container base image different from the host image affect the application?}
This is a question sometimes raised by HPC application developers and resumes to verify whether having different base images when deploying your application affects the performance even though the Linux kernel is being shared. The assumption is that the bundled packages and dependencies within the Image OS might affect performance. 

\begin{figure}[ht]
    \includegraphics[width=\textwidth]{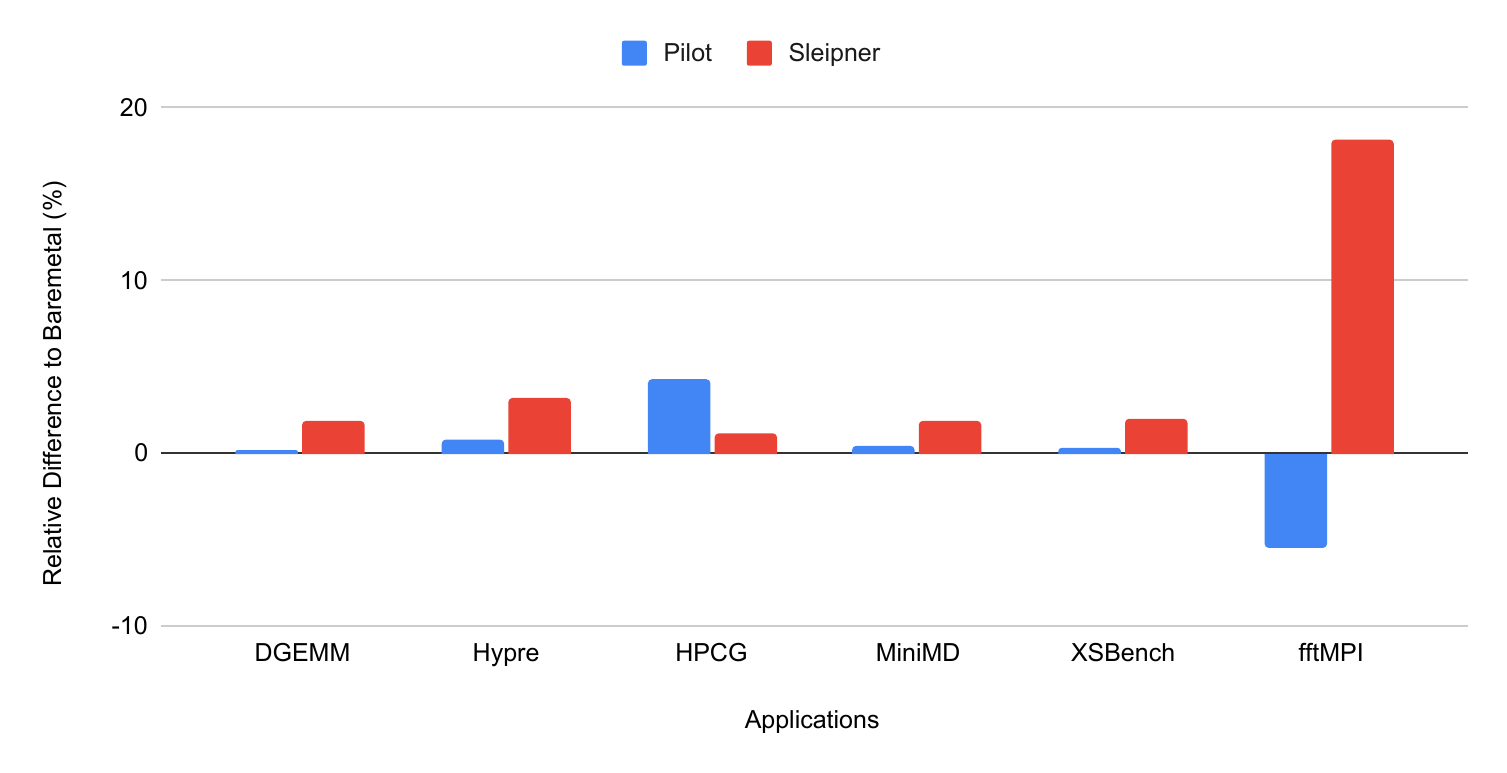}
    \caption{Performance comparison among different images on both Nvidia Grace platform and ARM Ampere Altra.}
    \label{fig:comparison}
\end{figure}

The experiment design for this question involved creating a different image, with the same software stack (compilers, and application libraries), but with a different image OS than the Host OS. We design an experiment through the usage of three applications (DGEMM, HPCG and Hypre) and in two systems (Sleipner and Pilot). Instead of using Rocky Linux and OpenSUSE, as was done in the previous Section, we used Debian Trixie. The difference in performance results can be seen in Figure \ref{fig:comparison}. In practice, we see that the performance impact is between nearly 0 to a bit over 4\%. In comparison to the results obtained in Q1 (where the image was roughly the same as the Host OS), we usually see a usual slowdown except for fftMPI, where apparently the layer of libraries provided by the Image OS was so different than the Host OS to the point of causing a 5\% increase in speed, while in Sleipner it created a 20\% slowdown. 

\subsection{How much a security-hardened rootless containers affect the performance of applications?}
The root privileges given inside containers is a usual point raised by system administrators in HPC clusters who would rather have their containers executed in rootless mode. Docker requires that at least its daemon executes with root privileges inside the system, while Podman is usually regarded as a rootless solution

That said, we strive to understand whether there is any impact of switching between different container engine privileges, and thus this third experiment was designed. We used our Cloudlab infrastructure, in which we have full control, to install and observe how the execution time of containers executed as root and rootless affects the overall performance in different applications. In Docker, we define a container as "root" or "rootless" depending on where the docker daemon would run. For Podman, this definition applied depending on which user would start the container (i.e., the "root" user starting the container would be considered root). 

\begin{figure}[ht]
    \includegraphics[width=\textwidth]{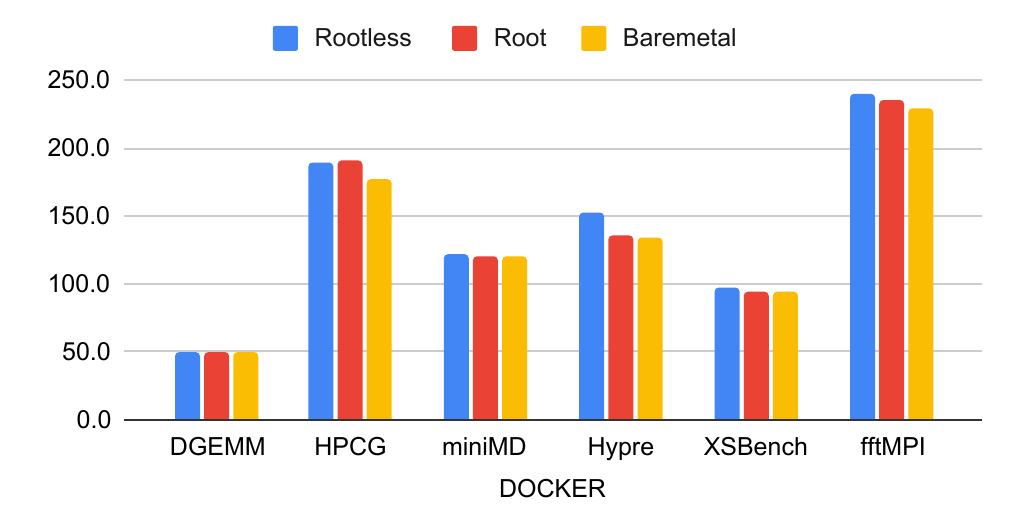}
    \caption{The performance comparison of rootless and root container execution modes on the Cloudlab platform in Docker and Podman, respectively.}
    \label{fig:rootless-docker}
\end{figure}

\begin{figure}[ht]
    \includegraphics[width=\textwidth]{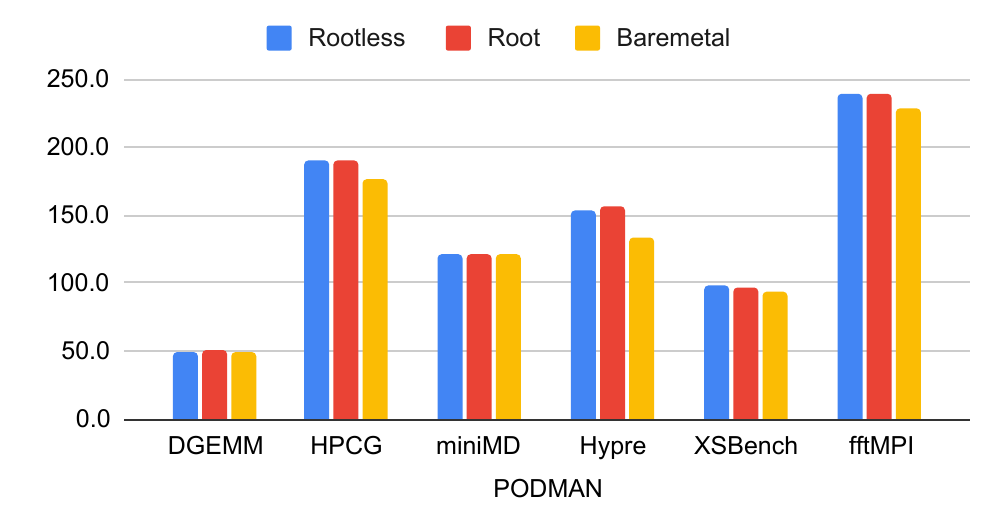}
    \caption{The performance comparison of rootless and root container execution modes on the Cloudlab platform in Docker and Podman, respectively.}
    \label{fig:rootless-podman}
\end{figure}

Performance results for each case can be seen in Figures~\ref{fig:rootless-docker} and ~\ref{fig:rootless-podman}. In practice, while running an application in bare metal mode is slightly better than using containers, there is not any significant difference between executing different privilege modes. 

\section{Related Works}

In virtual machines (VM), virtualization overhead was a concern in HPC applications. Cloud execution of tightly-coupled HPC workloads was previously shown to be several orders of magnitude slower than on a typical HPC cluster, and exhibited lower scalability on the cloud~\cite{jackson_performance_2010}, notably due to poor performance of MPI collective operations~\cite{chakthranont_exploring_2014}. The slower network interconnect in the cloud was found to be responsible for the poor performance of communication-intensive applications, this was identified to be partially due to the virtualized I/O to the network interface, that direct VM-level access could solve~\cite{exposito_performance_2013}. Regola et al.~\cite{regola_recommendations_2010} identifies that virtualization has a low overhead on computing performance. However, the performance problem mostly lies in I/O overhead. Exposito et al.~\cite{exposito_performance_2013} identify that virtualized access to the NIC is still a bottleneck, but observe a low overhead in shared-memory message passing. The authors propose message-passing plus multithreading as a scalable and cost-effective solution to run HPC applications on Cloud. 

On recent cloud platforms that use lightweight container-based virtualization, new optimizations are focused on improving the scalability of HPC applications. Milroy et al.~\cite{milroy_one_2022}~propose a multi-layer model to guide the porting of MPI-based HPC applications to Kubernetes using a declarative approach, along with allowing for automated MPI application lifecycle. They demonstrated strong scaling performance for a medium-scale HPC job, with 3000 MPI ranks. De Sensi~\cite{de_sensi_noise_2022} identify that due to the congestion-prone network on the cloud, in contrast to networks on HPC systems, co-running jobs on the cloud can cause noise and interference impacts the performance of HPC applications running on the cloud.

The portability of using containers on cloud and HPC systems has also been explored. Sindi et al.~\cite{sindi2019using} proposed to leverage the portability of containers for resilience on HPC clusters. They leverage performance counter monitoring to predict the failure of the compute node and migrate a container to another node without application modification using the CRI-U tool. Younge et al.~\cite{younge2017tale} also studied container portability, focusing on migrating from laptops to clouds or HPC systems. Their study used the Singularity runtime and two HPC benchmarks, HPCG and IBM on a Cray system, to pinpoint the overhead and scalability with containers.
\section{Conclusions}
Recent development in lightweight OS-level virtualization provides a potential solution for running HPC applications on the cloud platform, a path that was not widely adopted in the past due to the high overhead of hypervisor VM-based virtualization. In this work, we focus on the impact of changing the major layers in a containerized environment when migrating from a dedicated HPC system to the cloud platform. On the three platforms, with six representative applications, we characterized the impact of container virtualization, changed host OS and kernel, and changes in the rootless and privileged container execution. Our results indicate less than 4\% container overhead in applications that are not memory intensive, but significantly higher overhead (8\%-10\%) in FFT, HPCG, and Hypre. We also show that changing between the container execution modes results in negligible performance differences in the six tested applications.

\section*{Acknowledgments}
This research is supported by the European Commission
under the Horizon project OpenCUBE (GA-101092984).


%
%
\bibliographystyle{splncs04}
\bibliography{main}
\end{document}